\documentclass[12]{WileyMSP-template}

\usepackage{xr-hyper}
\usepackage{mathptmx}

\usepackage{siunitx}
\usepackage[dvipsnames,svgnames,x11names,table]{xcolor}
\usepackage[colorlinks, citecolor = Royalblue, urlcolor = RoyalBlue]{hyperref}
\hypersetup{
    citecolor = RoyalBlue,
    colorlinks=true,
    linkcolor=RoyalBlue,
    filecolor=black,      
    urlcolor= black,
    pdftitle={Overleaf Example},
    pdfpagemode=FullScreen,
    }

\begin{document}

\title{Creation of Negatively Charged GeV and SnV centers in Nanodiamonds via Ion Implantation}
\maketitle

\justifying
\author{Selene Sachero$^1$,}
\author{Richard Waltrich$^1$,}
\author{Emilio Corte$^2$,} 
\author{Jens Fuhrmann$^1$,}
\author{Sviatoslav Ditalia Tchernij$^2$,}
\author{Fedor Jelezko$^{1,3}$,}
\author{Alexander Kubanek$^{1,3}$*}

\begin{affiliations}
\sloppy
$^1$ Institute for Quantum Optics, Ulm University, Albert-Einstein-Allee 11, 89081 Ulm, Germany\\
$^2$ Physics Department, University of Torino, via P. Giuria 1, 10125 Torino, Italy\\
$^3$ Center for Integrated Quantum Science and Technology (IQst), Ulm University, Albert-Einstein-Allee 11, 89081, Ulm, Germany\\
*Corresponding author: alexander.kubanek@uni-ulm.de

\end{affiliations}


\begin{abstract}
\sloppy
Solid state quantum emitters, in particular group-IV vacancy centers in diamond, are at the forefront of research in quantum technologies due to their unique optical and spin properties. Reduction of the diamond host size to the nanoscale enables new opportunities in terms of integration and scalability. However, creating optically coherent quantum emitters in nanodiamonds remains a major challenge. Here, we present the fabrication of germanium-
and tin- vacancy centers by means of ion implantation. We describe the fabrication process and present the optical properties of the created color centers. We achieve high purity single photon emission via resonant excitation and strong coherent drive of a SnV$^-$ center. The successful integration of heavier group-IV vacancy centers in nanodiamonds paves the way for further advances in fields like hybrid quantum photonics or sensing on the nanoscale.

\end{abstract}

\section{Introduction}
Coherent quantum emitters are key building blocks on the way
to establishing quantum-based networks \cite{Kimble2008,Wehner2018, Hucul2015,quantum_network} or quantum repeaters \cite{Repeater,Repeater2,Gregor_2023}. Ideally, the interface would be modular, functionally scalable and mass-producible. Integrating solid-state quantum emitters with photonic devices is an appealing approach to solve the challenge of implementing such an interface \cite{Lukin_2016, Lifetime_SiV}. One alternative approach is to employ a hybrid system \cite{Hybrid, Wan_2020} comprising of a nanodiamond (ND) as the color center host, which is then selectively positioned within, for instance, silicon-based photonic devices  \cite{Niklas_2024}, micro-cavities \cite{Johnson_2015, Becher_2015_cav, Feuchtmayr2023} or bullseye structures \cite{Richard_Bullseye}. 
NDs hosting color centers, are not only essential for application in quantum optics but they have attracted remarkable attention for bioimaging \cite{nontoxic,bioimaging} and quantum sensing on the nanoscale \cite{sensing, sensing_NV}. It is therefore of great interest to reliably incorporate color centers into NDs maintaining their good optical properties. 

\begin{figure}[t!]
\includegraphics[width=\textwidth]{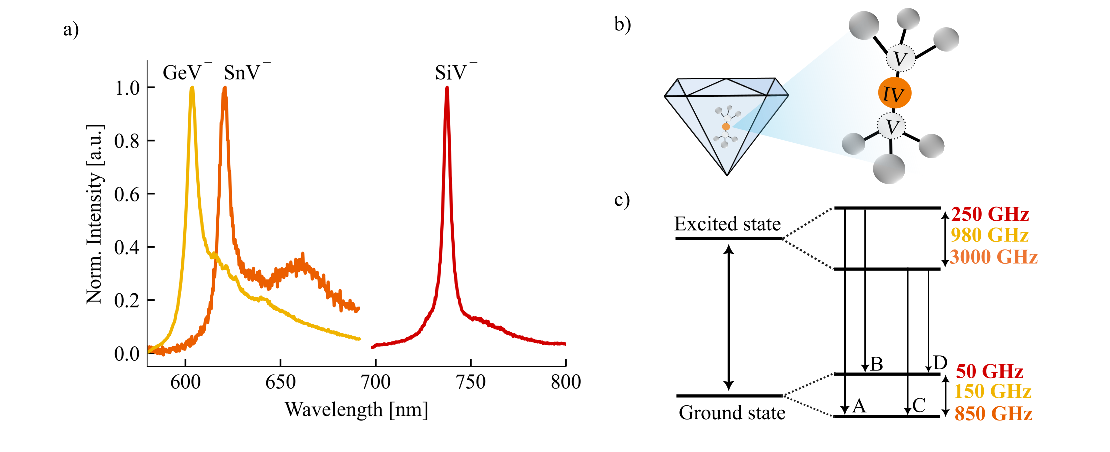}
\caption{\justifying \label{fig:1}Common properties of group-IV vacancy centers in diamond. a) Room temperature PL spectrum of the color centers showing their typical ZPL. b) Atomic structure of group-IV vacancy centers in the diamond lattice. The group-IV atom (IV, orange) lies in between two missing carbon atoms (V, white). c) Energy level structure within zero magnetic field and theoretical energies of the GS and ES. Red values for SiV$^-$, yellow values for GeV$^-$ and orange values for SnV$^-$.}
\end{figure}

As a coherent photon source, inversion-symmetric negatively charged group-IV vacancy (G4V) centers, in particular silicon vacancy (SiV$^-$), germanium vacancy (GeV$^-$) and tin vacancy (SnV$^-$) centers, are emerging candidates \cite{network_group_IV, network_building, Jantzen2016}. These emitters offer a high Debye Waller factor which results in a predominant emission into the zero-phonon line (ZPL) (figure \ref{fig:1}a)), crucial for spin-photon entanglement \cite{Germanium_2015, Becher_SiV_DB,gorlitz_2020}. The ZPL wavelengths for the SiV$^-$, GeV$^-$ and the SnV$^-$ centers are 737 nm, 602 nm and 620 nm respectively \cite{SiV_ZPL,GeV_ZPL, Iwasaki2017}. Additionally, their atomic structure, consisting of a group-IV element bonded to two adjacent missing carbon atoms, results in a $D_{3D}$ symmetry (figure \ref{fig:1}b)) which makes them robust against disturbances from external electrical fields \cite{Becher_2014,shallow_implantation,stark_effect}. The electronic structure of G4V centers is a spin 1/2 system which results in an electronic energy level scheme as shown in figure\ref{fig:1}c). Spin-orbit coupling leads to the formation of four orbital levels, two arise from the splitting of the ground state (GS) and two from the excited state (ES) \cite{Hyperfine_struct}. The resulting optically active transitions are labeled as A, B, C, D in figure \ref{fig:1}c) and can be observed only at cryogenic temperatures. Depending on the impurity atom the transition energies vary as well as the energy splitting of the GS and ES. The respective GS splittings for SiV$^-$, GeV$^-$ and SnV$^-$, at zero strain, are 50 GHz, 150 GHz and 850 GHz and the ES splittings are 250 GHz, 980 GHz and 3000 GHz (figure \ref{fig:1}c)) \cite{Becher_2014,Bhaskar2017, Iwasaki2017}.

Among G4V centers, the SiV$^-$ is the most studied \cite{Richard_2023, Rogers2014, Rogers_2019, Maletinsky_2023} and lot of previous work has been done to optimize the fabrication process both via ingrowth synthesis \cite{SiV_quantum_tec, Igor_2009, GeV_SiV_in_NDS} and ion implantation \cite{implantation_SiV, SiV_IonImplantation, Wang_2006}. In contrast, the deterministic creation of heavier G4V centers remains challenging. It is nevertheless desirable due to the good coherence properties of SnV$^-$ and GeV$^-$ centers at liquid helium temperature \cite{T2_tin, Kathi_2024}.

G4V centers are commonly fabricated via impurities incorporation during Chemical Vapor Deposition (CVD), high-pressure high-temperature (HPHT) synthesis, or via ion implantation \cite{PhysRevApplied.5.044010}. It has been demonstrated that the incorporation of Si and Ge atoms into nanodiamonds during CVD or HPHT growth process is possible due to their presence in the precursors utilized during the fabrication process \cite{Kondrin_2018, Sedov_2018}. Conversely, as the effectiveness of the synthesis diminishes with increasing atomic size \cite{atomic_size,PALYANOV2019769,Size_2} the most prevalent methodology for the generation of heavier G4V centers, such as the SnV$^-$ center, is ion implantation \cite{Slava}. 

Here, we demonstrate, for the first time, the fabrication of GeV$^-$ and SnV$^-$ centers in nanodiamonds via ion implantation and thermal annealing. The optical properties of the created color centers are investigated by means of resonant and off-resonant excitation at 4 K. The emission lines exhibit a narrow spectral distribution, centered on the expected values of the ZPL transition. High purity single photon emission and strong coherent optical driving are observed.

\section{Method}
\subsection{Fabrication process}

\begin{figure}[t!]
\centering
\includegraphics[width=\textwidth]{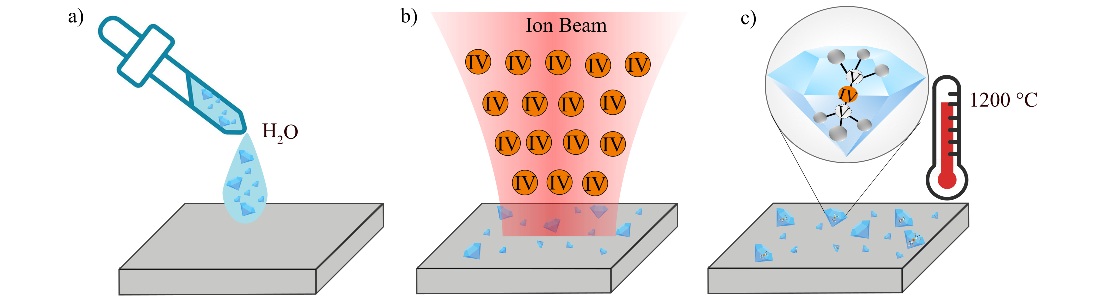}
\caption{\justifying\label{fig:sample_prep} Sketch of the sample preparation procedure.
a) The NDs were diluted in ultrapure water and a small droplet of 10 $\mu l$ was coated on the sapphire substrates.
b) The NDs were implanted with Sn and Ge ions (IV, orange). 
c) The implanted samples were annealed at 1200 $^\circ\text{C}$ to promote the formation of the color centers.}
\end{figure}

Commercially available size-selected NDs were diluted in ultrapure water and sonicated in an ultrasonic bath for 20 minutes to reduce agglomeration and finally coated onto two sapphire substrates with focused ion beam (FIB) milled markers (Figure \ref{fig:sample_prep}a)), which were previously cleaned in a triacid solution at 120 $^\circ\text{C}$ for three hours. The batch of NDs, known as Microdiamant MSY from Pureon Company with an average size of 180 nm was chosen \cite{Pureon_NDs}. This size allows integrability into micro-cavities without introducing scattering losses \cite{Gregor_2023, Florian} while also increasing the probability of success during the implantation process. 

\sloppy
Subsequently, the NDs on the two sapphire substrates were homogeneously implanted at the Laboratory of Ion Implantation of the University of Turin (LIUTo facility) \cite{Implanter_Turin}. The two samples were irradiated with a broad beam of 3 mm$^{2}$ with ${}^{74}Ge^{-}$ and ${}^{120}Sn^{-}$ ions at an energy of 57 keV and 60 keV and a fluence of $1 \: x \: 10^{14} \: ions/cm^{2}$ and $ 6\: x\: 10^{12} \:ions/cm^{2}$ respectively (figure \ref{fig:sample_prep}b)). Subsequently, to promote the formation of the color centers, the samples were annealed in vacuum (at pressure $ < 2 x 10^{-7}$ mbar) for one hour at 1200 $^\circ\text{C}$ (figure \ref{fig:sample_prep}c)). 

\subsection{Characterization}
\sloppy
The two samples were investigated with a home-built confocal microscope at cryogenic temperatures. The temperature on the NDs varied from (5-12) K. A 532 nm continuous-wave (CW) laser was employed for off-resonant excitation and to measure photoluminescence (PL) spectra. To excite resonantly, the Classic Dye CW Ring Laser from Sirah Lasertechnik was utilized. Both excitation lasers were focused onto the sample with an air objective with a 0.95 numerical aperture, which collected the fluorescence emitted from the color centers. A 50:50 beam cube split the fluorescence in two paths. 50{\%} of the signal was collected through a single mode fiber (SM), which acted as a pinhole and guided the fluorescence to a single-photon counting module. The remaining 50{\%} was directed to a spectrometer. In order to perform correlation measurements in Harbury-Brown and Twiss (HBT) configuration, both arms were connected with a SM fiber to two single-photon counters. As a correlation device, the Time Tagger Ultra module from Swabian Instruments GmbH was used. In order to select the collection fluorescence optical filters were placed before the beam splitter. 

Off-resonant excitation was used to confocally scan the samples, identify the position of the NDs and to acquire the PL spectrum of many color centers in order to estimate the average ZPL position. To confirm the success of the implantation process, we compared the total number of identified NDs with the number of NDs showing the presence of GeV$^-$ or SnV$^-$ centers. 

After identifying a ND containing a cluster of emitters we performed photoluminescence excitation spectroscopy (PLE) by tuning a narrow band laser across the C optical transition while detecting fluorescence in the phonon sideband. As a demonstration of the high purity single-photon emission, we measured, under resonant excitation, the second-order photon autocorrelation function, $g^{(2)}(t)$, and fitted the data with a two-level model function:

\begin{equation}
\label{eq_g2}
g^{2}(t) = 1 - A \cdot \exp (-|(t - t_0) / \tau|)
\end{equation}

We further coherently drove a single emitter and observed Rabi oscillations, which were fitted with the following equation: \cite{Steck, Rabi_equation, Coherent_oscillation}

\begin{equation}
I = 1 - \mathrm{e}^{-\frac{3\Gamma}{4} |t-t_{0}|} \left( \mathrm{cos}(\Omega |t-t_{0}|) + \frac{3\Gamma}{\Omega} \mathrm{sin}(\Omega |t-t_{0}|) \right)
\label{Rabi_eq}
\end{equation}

where $1/\Gamma$ is the dephasing time and $\Omega/2\pi$ represents the Rabi frequency.

\section{Results} 
\begin{figure*}[t]
\includegraphics[width=\textwidth]{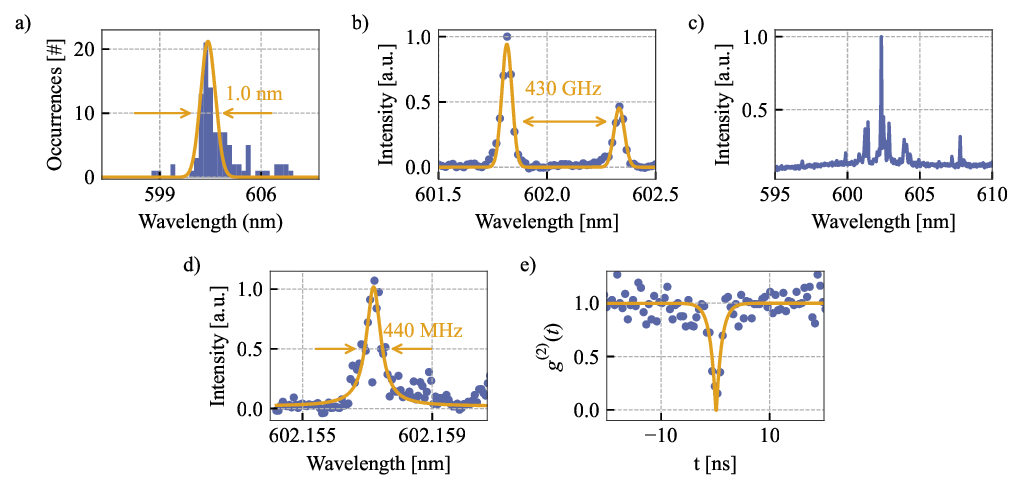}
\caption{\justifying \label{fig:4} Characterization of the optical properties of GeV$^{-}$ centers in NDs via off-resonant and resonant excitation at 4 K. a) Emission wavelength distribution centered at (602 $\pm$ 1) nm. b) Spectrum of an isolated emitter with a GS of (430 $\pm$ 20) GHz, the yellow line is a double Lorentzian fit. c) Spectrum of a cluster of emitters contained in a single ND. d) Average resonant PLE spectrum of a single GeV$^-$ center. The yellow line is the Lorentzian fit resulting in a linewidth of (440 $\pm$ 40) MHz. e) Resonant autocorrelation measurement without background correction, showing antibunching with single-photon purity of $g^{2}(0) = 0.15$.}
\end{figure*}

\subsection{\texorpdfstring{Spectroscopy of GeV\textsuperscript{-} centers}{Spectroscopy of GeV- centers}}
Thanks to the high implantation fluence ($10^{14} ions/cm^2$), the probability of finding NDs carrying GeV$^-$ centers is high, approximately 90$\%$. Meaning out of 100 NDs 90 carried at least one GeV$^-$ center. Confirming the success of the creation process, despite previously reported low value of creation yield for GeV centers \cite{Creation_yield}. However, the implantation process, with such a high fluence, resulted in the creation of a significant number of vacancies in the lattice (see section 1 in the SI ), leading to a high presence of nitrogen vacancy (NVs) centers and damages to the NDs lattice. Thus, in future, a lower implantation fluence on the order of $10^{12}$ $ions/cm^{2}$ is suggested. 

The histogram in figure \ref{fig:4}a) illustrates the inhomogeneous distributions of the ZPL for all measured GeV$^-$ centers. The data are distributed according to a normal distribution centered at $(602 \: \pm \:1)$ nm. In figure \ref{fig:4}b), we show that, despite the high implantation fluence ($10^{14}$ $ions/cm^{2}$), we were able to observe the fine structure of an isolated GeV$^-$ center. C and D transitions were fitted with a Lorentzian curve and the resulting wavelengths are $(601.82 \: \pm \:0.02) $ nm and $(602.33 \: \pm \:0.02)$ nm, respectively. This results in a GS splitting of (430 $\pm$ 20) GHz which indicates large strain in the surrounding ND lattice which originates from the damages induced by the high implantation fluence. We proceed with the optical characterization via resonant excitation. We selected a GeV$^-$ center from the cluster in figure \ref{fig:4}c). This emitter exhibited a resonant wavelength of (602.16 $\pm$ 0.02) and a linewidth  of (440 $\pm$ 40 ) MHz (figure \ref{fig:4}d)), which is four times broader than the lifetime limited linewidth reported for bulk diamond \cite{Bhaskar2017}. We address the remaining broadening to spectral diffusion which was observable during the PLE scans, and it is consistent with the persistent blinking fluorescence under CW resonant drive (see section 3 in the SI). To prove the quantum nature of the measured GeV$^-$ center CW resonant autocorrelation measurement was performed. The antibunching behavior results in a minimum value of $g^{2}(0) = 0.15 $. No background correction was performed on the data, so the high purity was a result of the resonant drive.

\begin{figure}[t]
\includegraphics[width=\textwidth]{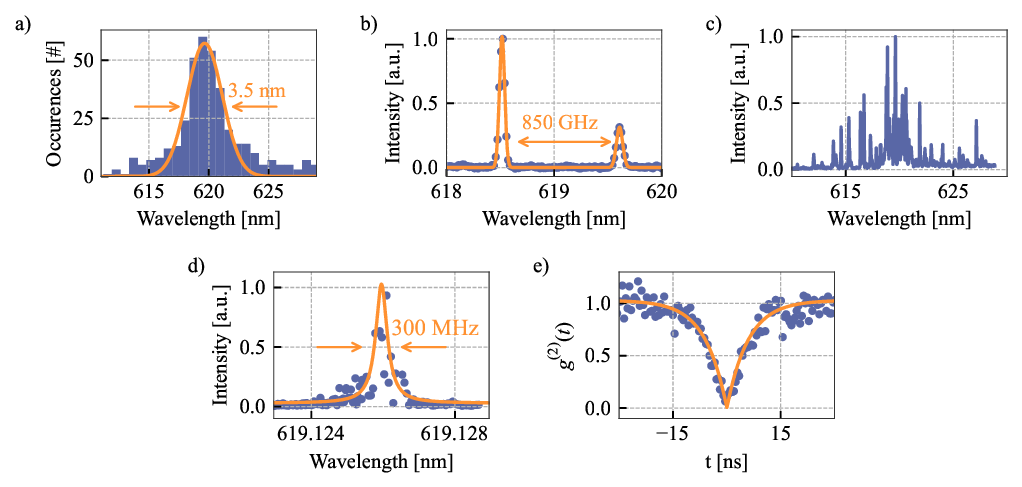}
\caption{\justifying Characterization of the optical properties of SnV$^{-}$ centers in NDs via off-resonant and resonant excitation at 4 K. a) Histogram of the distribution of the emission wavelength fitted with a Gaussian distribution centered at (620 $\pm$ 3) nm. b) PL spectrum of an emitter with a GS of (850 $\pm$ 20) GHz, the orange line represents a double Lorentzian fit. c) PL spectrum of a cluster of emitters contained in a single ND. d) Average resonant PLE spectrum of a single SnV$^-$ center. The orange line is the Lorentzian fit with a FHWM of (300 $\pm$ 30) MHz. e) Resonant autocorrelation measurement without background correction, showing strong antibunching with high single-photon purity of of $g^{2} (0) = 0.03$.}
\label{fig:5}
\end{figure}

\subsection{\texorpdfstring{Spectroscopy of SnV\textsuperscript{-} centers}{Spectroscopy of SnV- centers}} 

The larger the atomic size, the higher the probability of causing damage to the lattice \cite{chapter_implantation}. Therefore, a lower implantation fluence was chosen to create SnV$^-$ centers. However, the lower fluence reduced the probability of finding NDs containing at least one SnV$^-$ center compared to the GeV$^-$ sample. As a result, only 70\% of the identified NDs contained SnV$^-$ centers. Moreover, it was possible to identify SnV$^-$ single-photon emission at room temperature under off-resonant excitation (see section 2.1 in the SI).

The histogram in figure \ref{fig:5}a) illustrates the inhomogeneous distributions of the ZPL for all measured SnV$^-$ centers. The data are distributed according to a Gaussian distribution centered at $(620 \: \pm \:3)$ nm. Figure \ref{fig:5}b) presents an example of the PL spectrum of an isolated emitter. The fine structure of the emitter is evident; C and D transition lines were fitted with a Lorentzian and the resulted wavelengths are $(618.52 \: \pm \:0.02)$ nm and $(619.61 \: \pm \:0.02)$ nm, respectively. This results in a GS splitting of $(850 \: \pm \:20)$ GHz. The latter value is in accordance with the predicted theoretical value of 850 GHz \cite{Gali_2018} which indicates that the surrounding ND lattice has low strain. We proceed with the optical characterization via resonant excitation. We selected a single SnV$^{-}$ center from the cluster shown in figure \ref{fig:5}c). This emitter exhibited a resonant wavelength of $(619.12 \: \pm \:0.01)$ nm and a linewidth of $(300 \: \pm \:30)$ MHz as shown in figure \ref{fig:5}d). The measured linewidth was found to be significantly broadened compared to the lifetime-limited linewidth of 30 MHz reported for bulk diamond \cite{trusheim2020transform}. The broadening is attributed to spectral diffusion, which was observed during the PLE scans.
The broadening is consistent with the persistent blinking fluorescence observed under CW resonant drive (see section 2.2 in the SI). Despite the instability under resonant excitation, the emitter exhibited considerable brightness. Spectral diffusion can be attributed to a charge instability, which can be mitigated by oxygen annealing at 450 $^{\circ}{C}$ or by adding a CW 445 nm laser used to stabilize the charge \cite{Becher_chargeCicle}. The strong antibunching behavior (figure \ref{fig:5}e)), $g^{2} (0) = 0.03$ proved the high purity of the single-photon emission under resonant excitation. No background correction was performed on the data so the high purity was a result of the resonant drive.

\begin{figure}[t!]
\includegraphics[width=\textwidth]{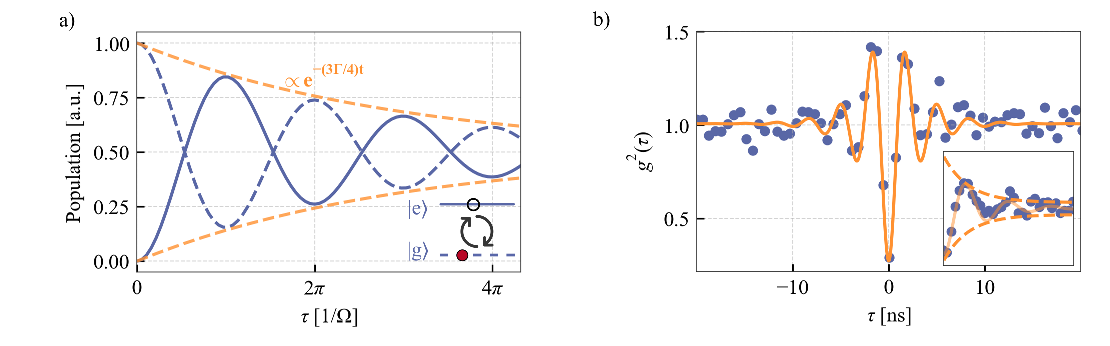}
\caption{\justifying \label{fig:Rabi} SnV$^-$ center coherent drive. a) Scheme for a two level system of the population oscillating in between the GS (dashed blue line) and the ES (solid blue line). On the right lower corner a two level system with an electron being excited from the GS to the ES. b) $g^{2}(\tau)$ showing Rabi oscillations. The data are fitted with equation \ref{Rabi_eq} (solid orange curve). Inset: zoom in of the $g^{2}(\tau)$ function and the envelop of the time evolution (orange dashed line).}
\end{figure}

\subsubsection{Coherent Drive}
When a two-level system is driven by a strong resonant electromagnetic field, it undergoes Rabi oscillations \cite{Rabi_1937}, the population periodically changes between the GS and the ES (figure \ref{fig:Rabi}a)). These oscillations decay exponentially over time, with the decay rate determined by the dephasing time of the system. This dephasing time characterizes how long the system maintains its coherence before interactions with the environment cause loss of phase information, leading to the damping of the Rabi oscillations. The oscillation frequency, known as the Rabi frequency, is determined by the laser intensity and the dipole moment of the transition. One method of observing Rabi oscillations is by measuring resonant autocorrelation functions, which describe the population dynamics.

To experimentally demonstrate the coherent drive, we measured a $g^{2}(\tau)$ under strong CW resonant drive laser. As shown in figure \ref{fig:Rabi}b) when a strong resonant laser was applied to the emitter with ZPL at 619.113 nm (present in the cluster in figure \ref{fig:5}c)) clear  Rabi oscillations were observed. These oscillations were detected at an excitation power of 200 nW (figure \ref{fig:Rabi}b)). By fitting the data with equation \ref{Rabi_eq}, we extracted the Rabi frequency ($\Omega/2\pi$) and the dephasing time. The Rabi frequency was determined to be ($300 \pm 5$) MHz an the dephasing time was $(1.93 \pm 0.19 )$ ns. In the inset of figure \ref{fig:Rabi}b) a zoom-in of the $g^{2}(\tau)$ measured at 200 nW is depicted. The dashed orange lines represent the envelope of the time evolution, which is described by an exponential decay proportional to the dephasing time. Additionally, we performed a second autocorrelation function at 75 nW (see section 2.3 in the SI). The estimated Rabi frequency and the dephasing time were ($165 \pm 5$) MHz and $(1.97 \pm 0.19 )$ ns respectively. As expected, the Rabi frequency increased with power \cite{rogers_all-optical_2014,Coherent_oscillation}, while the dephasing time remained constant at both excitation powers. This suggests the absence of excitation-induced dephasing, which can be partially explained by the insensitivity of SnV$^-$ centers to fluctuating electric fields within the diamond lattice \cite{Constant_T2}.

\section{Conclusion}
We demonstrated the formation of GeV$^-$ and SnV$^-$ centers in NDs by means of ion implantation and subsequent thermal annealing, and, for the first time, we showed coherent drive of a single SnV$^-$ center in a nanodiamond. 

The possibility of implanting such big atoms in NDs paves the way to creating coherent photon sources with the ability to control the size of the host diamond, by using size-selected NDs. 
In this work, a relatively high implantation fluence ($1x10^{14}$ $ions/cm^{2}$) was chosen for the GeV$^-$ sample, which led to a high presence of GeV$^-$ centers in the studied NDs. However, this process resulted in the creation of a significant number of vacancies that led to a high presence of NVs, which is attributed to the high implantation fluence, since prior to implantation no NVs were present in the NDs.

On the contrary, the SnV$^-$ sample was implanted with a lower fluence ($6x10^{12}$ $ions/cm^{2}$) which resulted in a lower probability, approximately 70\%, of finding SnV$^-$ centers in the characterized NDs. Reducing the implantation fluence decreased the number of vacancies (see section 1 in the SI), and, as a consequence, many NDs contained no NVs. Moreover, it was possible to identify single-photon emitters under off-resonant excitation at room temperature (see section 2.1 in the SI). 
In the future, by optimizing the implantation fluence, it will be possible to better control the number of emitters in a ND and the presence of NVs centers.

At cryogenic temperature, we observed that the majority of NDs showed clusters of multiple emitters, and only rarely isolated color centers were found. Nevertheless, it was possible to resonantly address single emitters as proved by the values of $g^{2}(0)$ well below 0.5. 

For the GeV$^-$ and SnV$^-$ center a linewidth of ($440 \: \pm \: 40$) MHz and ($300 \: \pm \: 30 $) MHz was measured, respectively. The broadening of the emission is attributed to spectral diffusion which can be mitigated, for SnV$^-$ , by oxygen annealing or by charge stabilization by applying an additional 445 nm CW laser \cite{Becher_chargeCicle}. Instead, the GeV$^-$ spectral diffusion can be due to damages originating from the implantation process. Thus, further improvement is required for GeV$^-$ implantation in order to achieve a better stability and coherent drive. Despite the broadening of the linewidths these two values are reported for the first time and prove that resonant excitation is possible. 

Furthermore, we demonstrated coherent and strong drive of single SnV$^-$ center, with a Rabi frequencies of ($165 \pm 5$) MHz at 75 nW and ($300 \pm 5$) MHz at 200 nW.

The presented results represent the initial stage in the creation of heavier G4V centers in nanodiamonds via ion implantation and further enhance the possibility of using color centers for quantum technologies. 

\section*{Acknowledgment}
The project was founded by the training LASer fabrication and ION implantation of DEFects as quantum emitters’ (LasIonDef) project funded by the European Research Council under the ‘Marie Skłodowska-Curie Innovative Training Networks’ program. We gratefully  acknowledge German Federal Ministry of Education and Research (BMBF) funding within the project QR.N (16KIS2208). We gratefully acknowledge further funding by the European Union Program QuantERA in the project SensExtreme (499192368).

Prof. Dr. Christine Kranz, Dr. Gregor Neusser and the Focused Ion Beam Center UUlm are acknowledged for their support in FIB milling. The author would like to thank Jens Fuhrmann for the annealing of the samples. We thank Pureon AG for providing ND samples. 

\section*{Data Availability Statement}
The data that support the findings of this study are available from the corresponding author upon reasonable request.

\bibliography{bibliography}
\end{document}